\documentclass[a4paper,11pt]{article}
\usepackage{pos}

\title{Star clusters as cosmic ray accelerators}

\author[a]{Stefano Gabici}

\affiliation[a]{Universit\'e Paris Cit\'e, CNRS, Astroparticule et Cosmologie, F-75013 Paris, France}

\emailAdd{gabici@apc.in2p3.fr}

\abstract{Massive stars blow powerful winds and eventually explode as supernovae. By doing so, they inject energy and momentum in the circumstellar medium, which is pushed away from the star and piles up to form a dense and expanding shell of gas. The effect is larger when many massive stars are grouped together in bound clusters or associations. Large cavities form around clusters as a result of the stellar feedback on the ambient medium. They are called superbubbles and are characterised by the presence of turbulent and supersonic gas motions. This makes star clusters ideal environments for particle acceleration, and potential contributors to the observed Galactic cosmic ray intensity.
The acceleration of particles at star clusters and in their surroundings may provide a major contribution to the observed CR flux. Moreover, it may explain the fine structures observed in the chemical composition of these particles, and possibly provide a solution to the puzzle of the origin of cosmic rays of energies in the PeV range and beyond.
}

\FullConference{%
  7th Heidelberg International Symposium on High-Energy Gamma-Ray Astronomy (Gamma2022)\\
  4-8 July 2022\\
  Barcelona, Spain\\}

\begin{document}
\maketitle

\section{Why star clusters as cosmic ray sources? I -- Cosmic ray composition}

Differences between the composition of cosmic rays (CRs) and of the Solar System can be used to characterise the acceleration sites of energetic cosmic particles \cite{wiedenbeckreview,lingenfelterreview}.
Refractory elements are overabundant in CRs \cite{meyer} when compared to volatiles, requiring a preferential acceleration of grain material over diffuse gas (see the pioneering works \cite{epstein,cesarsky}, where two distinct acceleration mechanisms for refractories were described).
A common feature of both scenarios \cite{epstein,cesarsky} is that refractories are accelerated at supernova remnant (SNR) shocks, which are believed to be the main sources of the bulk of Galactic CRs (see \cite{hillas,me} for critical reviews of this paradigm).
The scenario proposed by Epstein \cite{epstein}, and later consolidated by Meyer, Drury, and Ellison \cite{meyer, ellison}, provides a more viable explanation of the overabundance of CR refractories, and will be briefly described here.

Dust grains are characterised by a very large mass-to-charge ratio, and therefore by a very large rigidity.
The correspondingly large Larmor radius makes dust grains easily injected and accelerated at SNR shocks \cite{epstein,meyer,ellison}.
Once accelerated, dust grains are eroded by sputtering.
Sputtered ions (i.e. refractory elements) are expected to move at the same velocity of the parent grain, which is likely to be suprathermal.
Therefore, sputtered refractories produced in the upstream region of a SNR will be advected back to the shock and, due to their large Larmor radius, further accelerated.
It follows that the amount of interstellar refractories which undergo acceleration and eventually become CRs is {\it not} (or not only) determined by the injection mechanism into the acceleration process (as it is quite plausible that most if not all of the grains swept up by the shock undergo acceleration), but rather by the effectiveness of sputtering.
Meyer and co-workers \cite{meyer,ellison} provided a very convincing argument that this scenario can explain the observed overabundance of refractories in CRs.
They even went further to claim that the relative abundance of CR volatile elements might be explained by an acceleration mechanism whose efficiency scales as the element mass-over-charge ratio.
Recent simulations showed that diffusive shock acceleration might indeed satisfy this requirement \cite{damiano}.
The outcome of Meyer, Drury and Ellison's work \cite{meyer} is that the composition of CRs is controlled by {\it both} volatility and mass-to-charge ratio.

A limitation of these pioneering approaches is that they were based on near-Earth observations of CRs (the only ones available at that time) which are heavily affected by Solar modulation at low (GeV domain and below) particle energies.
As a consequence, measurements of the composition of CRs were quite uncertain.
This problem was solved when the two Voyager probes crossed the heliopause, providing us with a measurement of the pristine spectrum of interstellar CRs (or, in case some residual modulation still persists, with something very close to it) down to particle energies of few MeV/nucleon~\cite{voyager,me2}\footnote{A puzzling outcome of CR transport models in the Galaxy is that, in order to explain CR data at both low (Voyager) and high (AMS-02 and other detectors) energies, CR sources must inject {\it broken power laws} in momentum, rather than the simple power laws $\propto p^{-s}$ predicted by shock acceleration theory \cite{voyager,vincentreview}.
It is convenient to express the CR injection spectrum for the element $i$ as a function of the particle energy per nucleon, $\dot{q}_i(E) \propto \beta^{-1} p^{-s}$.
For energies above the break $s \sim 4.2-4.3$ for all (primary) elements, while below the break $s \sim 4$ for $i =$~H and He, but it is much harder for heavies (e.g. $s \sim 3.3$ for $i =$~O).
Moreover, the break energy is in the multi-GeV domain for H and at few hundreds MeV/nucleon for heavier elements \cite{vincent}.
The origin of these different behaviours is not understood, and calls for further investigations.}.
These new accurate measurements of CR composition can be used to characterise the phase of the interstellar medium out of which CRs are accelerated \cite{eichmann,vincent}.

To understand how this can be done, let us follow Meyer and collaborators \cite{meyer,ellison} and suppose that an element of mass number $A$ and charge number $Z$ is accelerated with an efficiency scaling as the mass-to-charge ratio to some power: $\propto (A/Z)^{\alpha}$. 
Simulations suggest, with quite some uncertainty, values of $\alpha = 1$ or 2 \cite{damiano}.
Consider then two extreme and idealised conditions characterising the gas reservoir out of which CRs are accelerated: an hot phase where all elements are fully ionised and a warm phase where all elements are singly ionised ($Z = 1$).
In the former case, all elements would be accelerated with roughly the same efficiency, as $A/Z \sim 2$, while in the latter the efficiency would scale with the mass number only, as $\propto A^{\alpha}$.
This illustrates how the ionisation state of the parent (pre acceleration) plasma impacts on the relative abundance of CR volatile elements.

The ionisation state is in turn set mostly by the (upstream) plasma temperature \cite{eichmann,vincent}, but a correction is needed if the acceleration operates at SNR shocks expanding in the warm phase of the interstellar medium.
This is because additional photoionisation of the ambient gas has to be taken into account.
It is due to UV and X-ray radiation from the supernova itself and from both the shocked ejecta and ambient medium (e.g. \cite{vincent} and references therein).
Once this correction is taken into account, the observed abundances of CR volatiles can be used to constrain the plasma temperature.
Tatischeff and collaborators \cite{vincent} found that data are best explained if the parent plasma is hot, with a temperature of the order of millions of degrees.
Such large temperatures are found inside superbubbles, i.e. the large cavities inflated in the interstellar medium due to the combined effect of stellar winds and recurrent supernova explosions in star clusters.
Moreover, the overabundance of CR refractories requires that such elements must be accelerated with an efficiency a factor of $\approx 10-50$ larger than that of volatiles.
To summarise, the results from Tatischeff and collaborators establish a clear link between volatile CRs and star clusters.
While the origin of refractories in CRs cannot be constrained in the absence of a clear knowledge of dust abundance and composition in the various phases of the interstellar medium, a scenario where {\it all} CRs (volatiles and refractories) are mainly accelerated in or around around star clusters is, to date, quite plausible (see discussion in \cite{vincent}).

While the scenario described above can reproduce the abundances of CR elements, it does not address the problem of {\it isotopic anomalies} which characterise energetic cosmic particles.
The most notable anomaly is the excess in the isotopic ratio $^{22}$Ne/$^{20}$Ne, which is equal to $\approx 0.3-0.4$ at CR sources, i.e. a factor of $\sim 4-5$ larger than in the Solar System, where the observed ratio is $\lesssim 0.1$\cite{binns,boschini}.
This prompted Casse and Paul \cite{casse} to suggest that a fraction of the CRs observed at the Earth must originate from chemically enriched wind material ejected by Wolf-Rayet (WR) stars.
As $^{22}$Ne is highly enhanced in WR stellar winds, some level of mixing of enriched (wind) and standard (Solar) material is required to reproduce the CR source isotopic ratio.

Three scenarios have been proposed to explain the $^{22}$Ne excess.
First, WR wind material can be accelerated by the SNR shock that forms after the star explodes \cite{meyer,prantzos}.
However, the model suffers from some fine tuning and, beside that, it is not clear whether WR stars eventually explode as supernovae or simply collapse into a black hole (see discussions and references in \cite{vincentreview,vincent}).
Second, the diffuse gas inside superbubbles may be chemically enriched due to the presence of WR stars in the parent cluster. Such material is then accelerated by SNR shocks as stars explode \cite{lingenfelterreview,higdon,binns}.
In this scenario, the composition of the material inside superbubble should be definitely non-solar, a thing which is not supported by X-ray observations of such objects (see e.g. \cite{jaskot}).
Finally, in the third and most promising scenario, enriched WR wind material is accelerated at the stellar wind termination shocks \cite{gupta,vincent}.
This latter possibility has been investigated quantitatively by Tatischeff and collaborators \cite{vincent}, who found that the $^{22}$Ne excess can be reproduced provided that $\approx 6$\% of the gas reservoir out of which CRs are accelerated is made of WR wind material.

Summarising, a link between volatile CRs and superbubbles (and therefore star clusters, which inflate such cavities) emerged from an accurate analysis of CR composition data.
Moreover, the most pronounced isotopic anomaly in the composition of CRs -- the excess in the $^{22}$Ne/$^{20}$Ne ratio -- can be explained by the acceleration of chemically enriched material at WR stellar wind termination shocks.
As massive stars are often found in clusters, it follows that also this isotopic anomaly is likely to originate in star clusters/superbubbles.

While it is not possible, at the moment, to decide whether also refractory CRs are originated at star clusters or not, the detection of 15 $^{60}$Fe nuclei in CRs might give us some hints \cite{60Fe}.
The $^{60}$Fe isotope is a short-lived radionuclide of half-life of $\sim$~2.6 Myr which is synthesised in core-collapse supernovae.
The fact that some energetic nuclei of this isotope reach the Earth means that their acceleration must have happened within the past few million years in the vicinity of the Solar system.
Assuming a {\it conventional} CR diffusion coefficient, it is possible to constrain the position of the acceleration site within a distance of $\approx$~1~kpc from the Sun (this is the displacement CRs would experience in a $^{60}$Fe half-life time).
A natural candidate for the acceleration of iron nuclei is the very nearby (less than 150 pc) Scorpius-Centaurus OB association \cite{60Fe}.
Recurrent supernova explosions took place in the association over the last few million years.
Such supernovae and their remnants might have been responsible for both the nucleosynthesis of $^{60}$Fe and its acceleration.
Moreover, the mechanical energy injected by stellar winds and supernovae in the Scorpious-Centaurus association is believed to be responsible for the formation of the Local Bubble within which the Solar system is located \cite{dieter}.
It is possible, even though far from being proven, that {\it all} CRs might come from star clusters/superbubbles, and we might even be located inside one of the accelerators of CRs!

\section{Why star clusters as cosmic ray sources? II -- Gamma-ray emission}

Enhanced gamma-ray emission is expected from CR sources, due to the decay of neutral pions produced when accelerated CRs interact with ambient matter.
In a pioneer paper, Montmerle~\cite{snob} used high energy gamma-ray observations to suggest that supernovae/OB associations (SNOBs) could be major sources of Galactic CRs.
This hypothesis remained for a long time much less popular than the mainstream scenario of CR acceleration at isolated SNRs, despite the fact that the number of associations of star clusters and superbubbles with GeV and TeV gamma-ray sources continued to grow slowly but steadily \cite{gammaclusters}.

The interest in stellar clusters as CR factories was revived when Aharonian and co-workers \cite{felixwinds}, after performing a morphological analysis of the high energy gamma-ray emission surrounding three of such objects, concluded that the emission had to be produced by CRs that had escaped the parent cluster after being accelerated there.
They also noticed that the gamma-ray spectrum of these objects extends to the multi-TeV domain without any sign of attenuation, implying that the spectrum  of the population of CRs responsible for the gamma-ray emission may reach the PeV energy domain and possibly beyond.
Finally, from the knowledge of the spatial distribution of the gas density, they could estimate the energy budget of CR in the regions, and concluded that the ensemble of {\it young}\footnote{The three objects they considered all have an age smaller than several million years.} star clusters in the Milky Way might provide a dominant contribution to the flux of the highest energy Galactic CRs.
As we will see in the next Section, young and old star cluster are expected to accelerate particles in a radically different way, and therefore it is important to increase the number of detections in gamma rays of objects of all ages.

In young clusters, the only source of mechanical energy are stellar winds. 
It is a well known fact that the integrated power of all stellar winds in the Galaxy is a sizeable but most likely minor fraction of the power injected by supernova explosions \cite{winds}.
Therefore, a scenario could be envisaged where the bulk of Galactic CRs is produced at SNR shocks, while stellar winds provide the energy to accelerate the particles that we observe at the high energy end of the Galactic CR spectrum.
A way to test this scenario would be to observe in gamma-rays very young clusters (with an age smaller than very few Myr), to avoid any contamination from supernova explosions.
An attempt to do so can be found in \cite{giada}.

In very recent times, the number of star clusters and superbubbles detected in GeV gamma-rays has further increased \cite{gammaclustersnew}.
In the TeV domain, observations exploiting the superior angular resolution of imaging atmospheric Cherenkov telescope arrays revealed a complex and quite extended morphology around the star cluster Westerlund 1 \cite{westerlund1}.
Further high angular resolution studies will certainly help in identifying the sites of particle acceleration and constrain the acceleration mechanism.
Remarkably, 2 out of the 12 sources detected by the LHAASO detector at sub-PeV gamma-ray energies might be associated with young massive star clusters \cite{lhaaso}, reinforcing belief that such objects might operate as the most extreme Galactic CR accelerators.

\section{Why star clusters as cosmic ray sources? III -- Ideal environment for particle acceleration}

In order to illustrate why the environment surrounding star clusters is ideal for particle acceleration it is convenient to analyse separately the case of young and old clusters.
Following \cite{varenna}, we call young clusters those where no supernova has exploded yet, and for this reason their mechanical energy output is uniquely due to stellar winds.
A cluster becomes old few million years after its formation, when supernovae begin to explode.

For young clusters, it is necessary to further distinguish between compact and loose ones \cite{gupta,varenna,thibault}.
In a compact cluster, stars are so close to each other that individual stellar outflows merge to forme a collective wind which is decelerated at an expanding and (in zero-order approximation) spherical Wind Termination Shock (WTS).
In loose clusters each star develops its own wind and WTS.
The theory of particle acceleration at WTSs was first developed for individual stars \cite{wts}, and much later generalised to the collective WTS that forms around compact clusters \cite{gupta,giovanni}.
Interestingly, while the WTSs around individual stars are invariably strong, the collective one around compact clusters are characterised by a finite value of the Mach number, which increases monotonically as the WTS expands \cite{gupta,varenna}.
This might have important observational consequences, as the theory of diffuse shock acceleration predicts that stronger shocks accelerate particles with a harder energy spectrum.
A comparison between theoretical predictions and observations (e.g. gamma-ray spectra) is needed in order to constrain models and validate our knowledge on particle acceleration at WTS.
Note that the vast literature devoted to the theoretical description of particle acceleration at the Solar WTS might be of great help, due to the similarity of the problem \cite{modulation}.

Let us now discuss old clusters.
The environment surrounding such objects is a low density cavity (a superbubble) inside which recurrent supernova explosions take place.
As more and more stars explode, the mechanical energy input from stellar winds decreases accordingly.
The interactions between stellar winds and SNR shocks makes superbubbles highly supersonic and turbulent environments.
Theoretical descriptions of the acceleration of particles in superbubbles have been developed over the years, mainly by Andrii Bykov and co-workers \cite{andrii}, by other authors \cite{others} and, very recently, by Vieu and collaborators \cite{thibault,thibaulthillas,thibaultnew}.
The results of such theoretical investigations show that the expected spectra of accelerated particles are far from being universal and, depending on the richness of the cluster, acceleration might operate in an intermittent way.
This implies that a case by case study of individual clusters is certainly not sufficient, and that observations of many of them are needed in order to figure out how acceleration proceeds in superbubbles.

Another important issue is that of the maximum energy a particle can achieve when it undergoes acceleration around a star cluster.
A model independent estimate is provided by the Hillas criterium~\cite{hillas}, that states that the maximum energy $E_{max}$ scales with the size of the accelerator $R$, the strength of the magnetic field there $B$, and the characteristic fluid speed of the system $u$, or: $E_{max} \sim R u B$.
For a young (old) cluster, $R$ is the radius of the WTS (the size of the superbubble), and $u$ is the wind termination velocity (the velocity of turbulent motions inside the superbubble).
The magnetic field strength $B$ is often unknown, but can be constrained to be at most a sizeable fraction of all available mechanical/thermal energy in the system.
When that is done one sees that, under most optimistic conditions, compact young stellar clusters can accelerate protons barely up to the PeV domain, while turbulent motions inside superbubbles are unable to do so \cite{varenna,giovanni,thibaulthillas}.

These estimates of $E_{max}$ are disappointing, as the acceleration of protons well beyond PeV energies is required to explain the smooth observed transition between Galactic and extragalactic CRs.
A way to overcome this problem is to speculate that $\gg$~PeV particles are accelerated at very fast SNR shocks expanding within the collective wind launched by a compact cluster \cite{thibaultnew} (the magnetic field is expected to be large in such an environment \cite{badmaev}).
The required shock speed would be of the order of $u \approx 30000$~km/s \cite{thibaultnew}, but this would in turn correspond to an explosion energy of $E_{SN} \approx 10^{52} (M_{ej}/M_{\odot}) (u/30000~{\rm km/s})^2$~erg, i.e. an extremely energetic explosion.
Very powerful explosions are very rare, and this might be a difficult problem to overcome \cite{pierre}.

\section{Conclusions and open questions}

After having discussed CR composition, radiative signatures of ongoing acceleration at star clusters, and acceleration mechanisms in such objects, it is tempting to advance the hypothesis that most, if not all Galactic CRs might be produced inside superbubbles.
Proving or disproving this hypothesis will be certainly one of the most crucial issues to be addressed in coming years.
Some of the main issues to be solved are: {\it i)} the puzzling spectral features observed in the composition of low energy CRs that are, to date, very poorly understood, {\it ii)} the (long standing!) difficulty of models in explaining how particles are accelerated up to the PeV domain and beyond, and {\it iii)} the identification of the acceleration site of Galactic CRs of different energies (isolated or clustered SNR? WTS? Turbulent bubbles?).

\textit{Acknowledgements:} SG thanks the organisers of the Symposium and acknowledges support from Agence Nationale de la Recherche (grant ANR-21-CE31-0028).
He also thanks V. Tatischeff, T. Vieu, and G. Peron for many enlightening discussions.

\end{document}